# Photoluminescence from nanocrystalline graphite monofluoride

Bei Wang[1], Justin R. Sparks[2], Humberto R. Gutierrez[1], Fujio Okino[3], Qingzhen Hao[1], Youjian Tang[1], Vincent H. Crespi[1], Jorge O. Sofo[1] and Jun Zhu[1]

[1]Department of Physics, The Pennsylvania State University, University Park, PA 16802, USA

[2]Department of Chemistry, The Pennsylvania State University, University Park, PA 16802, USA

[3]Department of Chemistry, Faculty of Textile Science and Technology, Shinshu University, 3-15-1 Tokido, Ueda 386-8567, Japan

We synthesize and study the structural and optical properties of nanocrystalline graphene monofluoride and graphite monofluoride, which are carbon-based wide bandgap materials. Using laser excitations 2.41 – 5.08 eV, we identify six emission modes of graphite monofluoride, spanning the visible spectrum from red to violet. The energy and linewidth of the modes point to defect-induced midgap states as the source of the photoemission. We discuss possible candidates. Our findings open the window to electro-optical applications of graphene fluoride.

Graphene's potential in optical applications remains largely untapped due to its gapless nature. Chemically modified graphene, such as graphene oxide (GO) or reduced GO have been shown to emit broad photoluminescence in the visible spectrum[1-4]. Although facile and effective, this approach does not produce the clean band gap desirable for many applications[3,5,6]. Crystalline, wide bandgap materials derived from graphene, such as graphane[7] and graphene fluoride[8,9], can potentially fulfill this requirement to produce a new class of optical materials. Density functional theory (DFT) calculations show a direct band gap of ~ 3.5 eV in graphene fluoride[8-10], which may increase to ~5 eV when quasiparticle corrections are included[7,10]. In addition to power electronics[11], graphene fluoride can potentially be used as a deep ultraviolet (UV) light emitter, similar to hexagonal boron nitride[12] or diamond[13]. Such devices may be effectively integrated with carbon electronics to achieve compact assembly and new functionalities.

In a previous study, we have synthesized graphite fluoride and graphene fluoride and confirmed its large band gap through electric transport measurements[9]. In this work, we report photoluminescence studies on nanocrystalline graphite monofluoride, using variable photon excitation energies at different temperatures. Six emission modes, from red to violet, are observed. These emissions likely originate from defect states residing in the band gap of CF.

We synthesize graphite monofluoride $(CF)_n$ from highly ordered pyrolytic graphite (HOPG) as previously described in Ref. 9 but with the reaction time extended to more than 100 hr. The resulting material appears white and flaky. As a reference, we note that SP-1 graphite fluorinated under identical conditions for 48 hr show a F/C ratio of 0.96 using chemical analysis (Laboratory for Organic Elemental Microanalysis, Kyoto University). X-ray diffraction measurements show multiple diffraction peaks and an inter-layer separation of 5.82 Å, in excellent agreement with existing data on $(CF)_n$[14] (See Fig. S1 in the supplementary material at [URL will be inserted by AIP]).

Thin sheets of $(CF)_n$ are mechanically exfoliated onto a transmission electron microscope (TEM) grid (Cu with lacey carbon support) and imaged in a JEOL-2010F microscope. The sheets vary in thickness and roughness, as reported previously in Ref. 9. Very thin sheets occasionally show step-like structures, consistent with the layered structure of $(CF)_n$. An example is given in Fig. 1(a). Electron diffraction (ED) of the circled area 1(0.25 $\mu m^2$) on a single terrace shows diffraction spots up to the 7$^{th}$ order and clear six-fold symmetry (Fig. 1(b)). We estimate the lateral size of a nanocrystalline domain to be $L_a \sim 20$nm from the size of the 1$^{st}$ order spots, taking into account instrument broadening calibrated by a reference graphene sample. Although area 1 encompasses many such nanocrystalline domains, their lattice orientations appear to align, as indicated by the well-defined spot locations. This situation changes as the electron beam is moved to a step edge of two terraces (area 2 in Fig. 1(a)). The ED pattern shown in Fig. 1(c) now consists of two sets of rotated patterns. As the electron beam includes more layers, ring-like structures appear (Fig. 1(d)), indicating higher stacking disorder. This observation is consistent with the very weak interlayer coupling of $(CF)_n$. Diffraction spots and rings collected from various locations on suspended CF sheets yield a lattice constant $d_{100}$ of (2.21 ± 0.03) Å, corresponding to an expansion of 4 ± 1 % compared to 2.13 Å in graphene. These results are in excellent agreement with the previously reported lattice constant of $(CF)_n$[15] and our DFT calculations of a single layer of graphene monofluoride, described below. The above structural characterizations indicate nanocrystalline graphene monofluoride to be the dominant building block of the synthesized compound.

We perform photoluminescence (PL) study on bulk $(CF)_n$ using four excitation laser lines ranging from green to deep UV (514, 488, 364, 244 nm). PL spectra are collected at room temperature, 85 K and 5 K using two different gratings: an 1800 groove/mm grating from 1.2 to 3 eV, and a 3600 groove/mm grating from 2.5 to 5 eV. Raw spectra are corrected by the efficiency of the gratings, which is calibrated using a halogen-deuterium balanced white light source. Eight PL spectra (a) – (h) are shown in Fig. 2. Spectra (a) and (b) are normalized to form a continuous trace from 1.5 – 3.4 eV based on overlapping data. Others are normalized to their highest intensities. The excitation energy of each spectrum, $E_{ex}$, is labeled in the unit of eV.

The emission profile we observed depends on both the excitation energy and sample temperature. Spectrum (c) obtained with $E_{ex}$ = 2.41 eV (514 nm) at room temperature shows a broad emission profile from 1.5 – 2.4 eV. On top of the background, four emission modes with roughly equal strength are visible. These modes are further resolved upon cooling. Spectra (e) – (g) are taken at 85 K with $E_{ex}$ = 2.41 (514 nm), 2.54 (488 nm), and 3.41 eV (364 nm) respectively. Each spectrum shows the typical emission profile of that $E_{ex}$, i.e. it is insensitive to the location of the laser spot on the sample. The four emission modes identified in spectrum (c) persist in (e) – (g) although their relative strength changes drastically. These modes, labeled $I_1$ – $I_4$, are marked by dashed lines and centered approximately at 1.70, 1.94, 2.07 and 2.25 eV. A slight upshift (< 0.02 eV) is sometimes observed with increasing $E_{ex}$. In the upper inset of Fig. 2, we show a tentative fitting to spectrum (e) using Lorentzian lineshapes. This fitting requires three additional modes, although their signatures in the spectra are very subtle. The full width at half maximum (FWHM) of $I_1$ – $I_4$ ranges from 0.08 – 0.17 eV. These values are much larger than the thermal broadening $k_BT$ = 8 meV at 85 K, suggesting the presence of defect-induced broadening. As $E_{ex}$ increases from 2.41 eV to 3.41 eV in spectra (e) – (g), higher-energy emissions gain strength. For $E_{ex}$ = 2.41 and 2.54 eV, the green mode at $I_4$ = 2.25 eV is the strongest. As $E_{ex}$ increases to 3.41 eV, a new mode centered at 2.8 eV (violet) begins to dominate the photoemission. This dominance is best seen in spectra (a) and (b). Upon cooling to 5 K (spectrum (h)), this 2.8 eV peak splits into two modes: $I_5$ at 2.67 eV and $I_6$ at 2.79 eV, as marked in the figure. The FWHM's of $I_5$ and $I_6$, 0.07 eV and 0.14 eV respectively, are similar to those of $I_1$ – $I_4$ and again suggest extrinsic broadening. As $E_{ex}$ increases further to 5.08eV, the main emission upshifts to 2.94 eV and acquires an asymmetric high-energy tail as shown in spectrum (d), with no other major modes. In particular, the band-edge emission of $(CF)_n$, which is expected to lie in the range of 3.5 – 5 eV, is absent[7,8]. The large width of $I_1 - I_6$, together with their sub-gap energies, suggests that they most likely originate from midgap states produced by defects. The absence of the band-edge emission may be due to either insufficiently large $E_{ex}$, or quenching from defect emission, as observed in defect-rich ZnO nanowires[16].

We made a critical observation of two UV Raman active modes in graphene monofluoride. As shown in the lower inset of Fig. 2, a sharp peak appears at 1270 (32) cm$^{-1}$, accompanied by a weaker peak at 1345 (43) cm$^{-1}$. Density functional theory calculations in both the frozen phonon and linear response approximations indentify the higher-frequency mode as $A_{1g}$ mode with out-of-plane motions of F against C and the lower-frequency mode as a two-fold degenerate in-plane $E_g$ vibration, and the linear response results are consistent with the $E_g$ mode having the stronger Raman response (See supplementary material at [URL will be inserted by AIP] for details). These sharp Raman features provide further evidence to the crystallinity of $(CF)_n$[17]. Their appearance upon UV excitation suggests resonant pumping[17], and implies that the band gap of $(CF)_n$ is very close to 5.08 eV.

We suspect that all defect states originate within the graphene monofluoride plane, given the large inter-layer separation of $(CF)_n$. Primary candidates are isolated or clustered unfluorinated carbon atoms and unsaturated bonds at nanocrystalline domain boundaries[18]. In addition, hexagon $sp^2$ carbon rings may be present, as indicated by the appearance of the *D* band in Raman obtained with 514 nm excitation[19] (See Fig. S2 in supplementary material at [URL will be inserted by AIP]). The ladder of photoluminescent states suggests a similar hierarchy of well-defined defect structures. The precise nature of the defects, as well as how to tailor and eliminate them to control the emission of graphene monofluoride, will be the subject of future studies.

In conclusion, we synthesize and report the optical properties of nanocrystalline graphite and graphene monofluoride. Graphite monofluoride exhibits six photoluminescent modes from red to violet. These modes likely originate from midgap states induced by defects in the plane of graphene monofluoride. Our experiments point to the possibility of using graphene fluoride in electro-optical applications.

We thank John Badding for providing access to his UV optics and Shih-Ho Cheng, Rongrui He and Benjamin Cooley for technical assistance. This work is supported by NSF NIRT grant

No. ECS-0609243. F. O. acknowledges the support of Regional Innovation Cluster Program of Nagano, granted by MEXT, Japan. The authors acknowledge use of facilities at the PSU site of NSF NNIN and the Penn State Materials Characterization Lab.

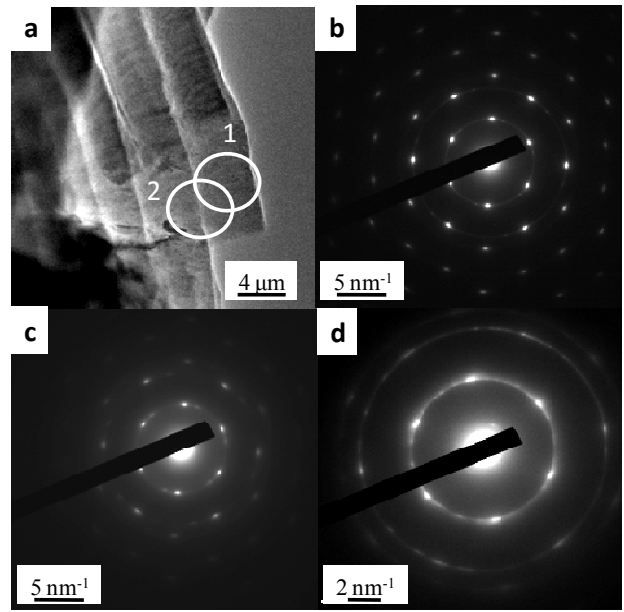

Figure 1. (a) Bright field TEM image of a thin sheet of graphene monofluoride suspended on a TEM grid. The contrast comes from thickness variation. (b) and ( c) Electron diffraction pattern from circle "1" and "2" in (a) respectively. (d) A typical electron diffraction pattern from thick sheets of graphite monofluoride.

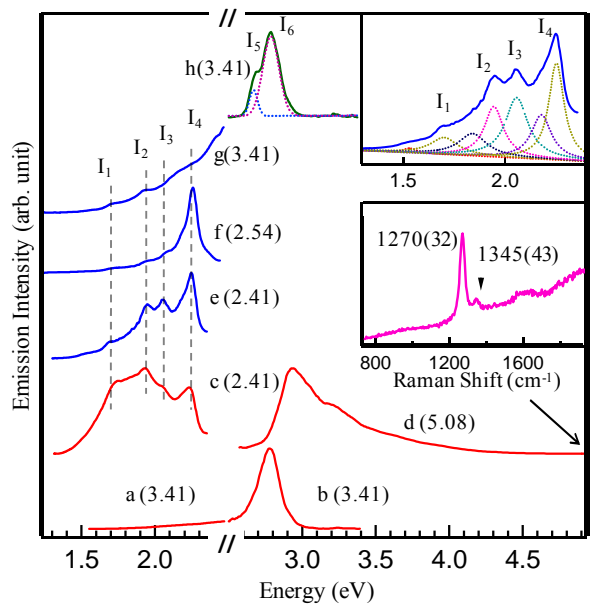

Figure 2. (color online) Photoluminescence from bulk (CF)$_n$, taken at 295K(a–d, red lines), 85K (e–g, blue lines) and 5K (h, green line). The excitation energies are labeled in units of eV. The dashed lines mark emission modes at 1.72, 1.94, 2.07, 2.25, 2.67 and 2.79eV. The top inset shows a tentative fitting of spectrum (e). The bottom inset shows a UV Raman spectrum of (CF)$_n$ observed at the high energy tail of spectrum (d) with two identified modes marked in the figure.

# Photoluminescence from nanocrystalline graphite monofluoride (supplementary information)

Bei Wang[1], Justin R. Sparks[2], Humberto R. Gutierrez[1], Fujio Okino[3], Qingzhen Hao[1], Youjian Tang[1], Vincent H. Crespi[1], Jorge O. Sofo[1] and Jun Zhu[1]

[1]Department of Physics, The Pennsylvania State University, University Park, PA 16802, USA

[2]Department of Chemistry, The Pennsylvania State University, University Park, PA 16802, USA

[3]Department of Chemistry, Faculty of Textile Science and Technology, Shinshu University, 3-15-1 Tokido, Ueda 386-8567, Japan

## X-ray diffraction (XRD) measurements

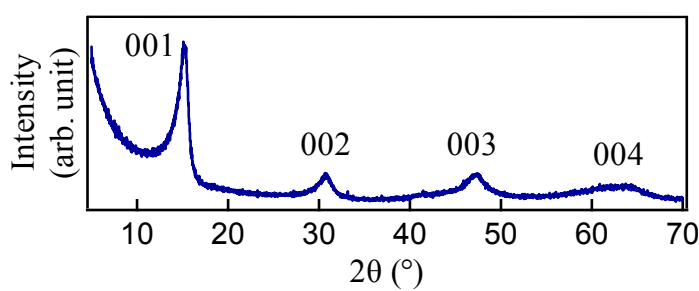

Figure S1. X-ray diffraction spectrum of fluorinated HOPG showing inter-plane diffraction peaks (001) – (004) as indicated in the figure. We extract an inter-layer separation of 5.82 Å from the peak positions. The crystalline size along the c-axis $L_c$ is estimated to be 6 nm from the width of the (001) peak. The in-plane diffraction peaks are absent due to the geometry of the setup.

## Raman spectroscopy using 514 nm laser excitation

Fig. S2 shows a Raman spectrum of $(CF)_n$ excited by a 514 nm laser line. The signature modes of $sp^2$ carbon, i.e., the $D$ mode at 1330 cm$^{-1}$, the $G$ mode at 1550 cm$^{-1}$ and the $D'$ mode at 1592 cm$^{-1}$ are present. In particular, the appearance of the $D$ mode suggests the presence of unfluorinated carbon rings[1]. These Raman modes are likely to be resonantly enhanced due to transitions to defect states at this excitation wavelength [2-4].

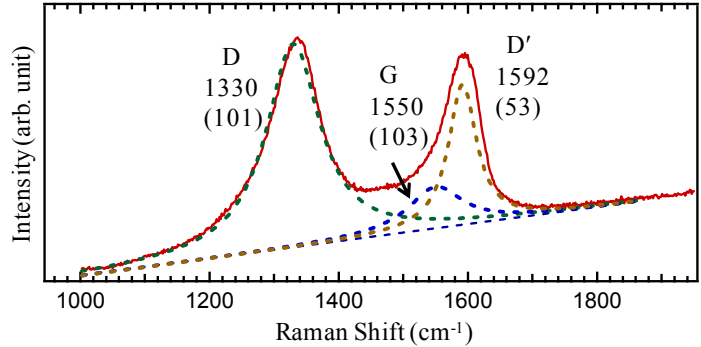

Figure S2. Raman spectrum of (CF)$_n$ using 514 nm laser excitation showing Raman modes from $sp^2$ carbon bonds.

**Density functional calculations of Raman response of graphene monofluoride**

Calculations using the projector augmented wave (PAW) method in both the local density (LDA) and generalized gradient approximations (GGA-PBE) were performed to identify the Raman-active modes of the CF samples shown in the inset to Fig. 2. Frozen phonon calculations used a plane wave cutoff of 400 eV and 24×24×1 k-points in a Monkhorst-Pack grid[5] using the Vienna Ab-initio Simulation Package (VASP)[6-9]. Linear response calculations (DFPT)[10] use a cutoff of 940 eV and a 6×6×2 grid, calculated with a plane-wave basis within GGA-PBE using the CASTEP code[11]. The in-plane lattice constant was relaxed in both cases. Within the linear response calculation, the core electrons are treated with norm-conserving pseudopotentials in a reciprocal space representation using core radii of 1.6, 1.5 and 1.6 Bohr for the *s*, *p*, and *d* channels of F and 1.0, 1.4 and 1.4 Bohr for the *s*, *p*, and *d* channels of C. The ionic displacement was converged to less than 0.0005 Å; the self-consistent electronic minimization was converged to an eigenenergy difference less than $0.8×10^{-10}$ eV for 3 iterations; and phonon modes at Γ were converged to a tolerance of 0.00001 eV/Å$^2$.

Table I summarizes the results of experiments and calculations using different methods. All methods consistently identify the 1270 (32) cm$^{-1}$ and 1345 (43) cm$^{-1}$ modes observed in the UV Raman spectra as the high-frequency $E_g$ and $A_{1g}$ modes respectively. Slight differences between different methods are correlated with differences in the relaxed in-plane lattice constants, in that longer C-C bonds produce slightly lower frequencies. Results for each method are numerically converged to approximately 10 – 20 cm$^{-1}$ precision. For simplicity, the inter-planar distance was taken large enough (20 Å) to produce isolated sheets. Inter-planar interactions between the weakly coupled CF sheets in experimental samples will shift the frequencies slightly. Most likely, interlayer interactions slightly stiffen the modes, thus accounting for the slight underestimate in the calculated frequencies. In addition to frequencies, linear response calculations provide direct access to the Raman tensor and hence the intensity of the Raman modes. Calculations show the $E_g$ mode at 1270 (32) cm$^{-1}$ to be the strongest, in agreement with experiments.

Table I. Summary of experimental results and density functional theory calculations for Raman and IR active modes in graphene monofluoride.

| Symmetry/Activity | Frequency (cm$^{-1}$) | Intensity (Å$^4$) | Method |
|---|---|---|---|
| $A_{1g}$ Raman | 1328 | – | DFT-LDA |
| | 1287 | – | DFT-GGA |
| | 1314 | 60 | DFPT-GGA |
| | **1345** | | Experiment |
| $E_g$ Raman 2-fold degenerate | 1242 | – | DFT-LDA |
| | 1187 | – | DFT-GGA |
| | 1161 | 117 | DFPT-GGA |
| | **1270** | | Experiment |
| IR Active | 1238 | – | DFT-LDA |
| | 1186 | – | DFT-GGA |
| | 1202 | | DFPT-GGA |
| | **1204**† | | Experiment |
| $A_{1g}$ Raman | 692 | – | DFT-LDA |
| | 670 | – | DFT-GGA |
| | 684 | 5 | DFPT-GGA |
| IR Active 2-fold degenerate | 292 | – | DFT-LDA |
| | 309 | – | DFT-GGA |
| | 333 | | DFPT-GGA |
| $E_g$ Raman 2-fold degenerate | 250 | – | DFT-LDA |
| | 251 | – | DFT-GGA |
| | 279 | 0.5 | DFPT-GGA |
| In-plane lattice constant: | 2.56 Å | | DFT-LDA |
| | 2.61 Å | | DFT-GGA |
| | 2.61 Å | | DFPT-GGA |
| | **2.57 Å** | | Experiment |

†Fourier transform infrared absorption measurements of (CF)$_n$ show a sharp peak at 1204 cm$^{-1}$ (unpublished).